\documentclass[11pt]{article}
\usepackage[margin=25mm]{geometry}
\usepackage{amsmath,amssymb,bm,graphicx}
\usepackage[numbers,sort&compress]{natbib}
\usepackage[colorlinks=true,allcolors=blue]{hyperref}
\usepackage{pdfpages}
\title{A disordered colloidal layer stores, rewrites and orders hidden memories}
\author{Vladimir A. Baulin\\[2pt]
\small Synthetix Institute, Keizersgracht 241, 1016 EA Amsterdam, The Netherlands\\
\small \href{mailto:vbaulin@synthetix.institute}{vbaulin@synthetix.institute}}
\date{}

\begin{document}
\maketitle

\begin{abstract}
\noindent
Information written into matter is usually carried by structure and kept
as a single record that bulk measurements can detect. Brownian simulations
show that a caged layer of colloidal ellipsoids at a liquid interface stores
many patterns of particle orientations, keeps each after the writing field is
removed and looks isotropic under a polarizer. Capillary bonds favour
orientations that are mirror images about the line joining two particles.
Because the bonds are fixed and disordered, the layer cannot forget by
rotating as a whole, and frustrated loops of bonds give it many metastable
patterns, whose number grows with the size of the layer. Lowering the
coupling during a write replaces a stored pattern, and successive writes
coexist as traces graded by age, so a single image shows the recent writes
and their order. The layer is a model for an optical memory that is
hidden, rewritable and records its own history.
\end{abstract}

Optical labels and exposure indicators share one limitation: what is written
into them is either fixed at manufacture or visible to anyone with the right
filter. A patch of material that light could write, that kept the information
without power, that looked uniform under a polarizer and that could be
rewritten would hide information in plain sight. If it also recorded which
exposures it had received and in what order, it would be a sealed log of its
own history, able to answer a sequence of optical challenges rather than a
single one. Electronic tags provide these functions with a chip and a power
source. Disordered optical media provide keys that are difficult to copy,
because their response is set by a random microstructure~\cite{pappu_physical_2002},
but that response is fixed once the medium is made. What is missing is a
material that stores several patterns in the same constituents while every
constituent keeps fluctuating thermally, and that develops no net order that
a bulk measurement could detect.

Matter can remember its history in many forms~\cite{keim_memory_2019}.
A sheared suspension retains several earlier strain amplitudes in the
arrangement of its particles~\cite{keim_generic_2011,paulsen_multiple_2014},
mechanical metamaterials can count and store driving cycles in the states of
bistable elements~\cite{chen_mechanical_memory_2021,kwakernaak_counting_2023},
and frustrated colloidal arrays have been written by setting the occupancies
of double wells~\cite{ortiz_ambriz_ice_2016}. In these systems the record is
carried by the structure: particles move, contacts change or elements snap.
In disordered Ising systems, the order in which athermal fields are applied
can determine the final microstate~\cite{croce_multi-field_2026}. A record of
a sequence in a thermal material whose particles stay in place would have to
be carried by internal degrees of freedom, and orientations are the simplest
of these. They change under a torque without displacing the particle. They
also diffuse, however, and ordinary alignment between neighbours drives the
layer towards one common direction, which erases any spatial pattern of
orientations.

A spatial pattern can be stabilized by interactions that refer to the
directions of the bonds between particles. When a bond prefers a pair of
orientations tied to the line joining the two particles, a disordered
arrangement gives each particle its own set of preferred directions, and the
preferences of different bonds conflict around closed loops. The layer then
has many metastable patterns, selected by the geometry rather than built into
the couplings. The situation resembles associative memory in networks of
interacting units~\cite{hopfield_neural_1982}, whose number of stored patterns grows with the number of units~\cite{amit_storing_1985}, and in which patterns written in
succession form a palimpsest: recent patterns are recalled, and older ones
fade as new ones are added~\cite{nadal_palimpsests_1986,parisi_forgets_1986}.
In a material the patterns are not stored in adjustable couplings but in the
state of the particles, and the question is whether the dynamics of the
particles alone produce such a graded record. Colloidal ellipsoids at a fluid
interface interact through the deformation of the meniscus, which has exactly
this dependence on bond direction~\cite{loudet_capillary_2005,fournier_anisotropic_2002}.

Here Brownian simulations show that a caged layer of these particles is that
material. Its memory is a pattern of particle orientations that the
capillary bonds keep after the writing field is removed, while the net
alignment stays near \(0.1\), so that a polarizer sees an isotropic layer. One
arrangement of particles supports many such patterns: each of 48 patterns
written separately into a layer of 2,304 particles is identified in every
final configuration, and the number that can be told apart grows with the
size of the layer. Lowering the coupling while a field acts erases a stored
pattern and writes a new one, and patterns written in succession remain as
traces graded by age, so one image of the particle axes shows the recent
writes and their order. The metastable patterns of the layer, fixed by the
random arrangement of its particles, are at once the states in which
information is written and the key with which it is read.

\section*{Capillary bonds tie orientations to the bond directions}

A bond can act on orientations in two ways. One contribution favours
parallel neighbours and depends on the difference between their angles.
The other favours angles related by reflection across an axis set by the
bond. Let \(z_i=e^{\mathrm{i}\psi_i}\) denote the internal orientation at
site \(i\), where \(\psi_i\) is the physical angle for a polar moment and
twice the physical angle for an apolar axis. The pair energy of this class
of interactions is
\begin{equation}
 U=-\sum_{(ij)}\left[A_{ij}\operatorname{Re}(\bar z_i z_j)
       +B_{ij}\operatorname{Re}(e^{-\mathrm{i}\chi_{ij}}z_i z_j)\right],
 \label{eq:class}
\end{equation}
where the sum counts each pair once, the bar denotes complex conjugation,
\(A_{ij}\) and \(B_{ij}\) are coupling energies and \(\chi_{ij}\) is the
angular preference of the bond. In terms of angles,
\begin{equation*}
 U=-\sum_{(ij)}\left[A_{ij}\cos(\psi_i-\psi_j)
       +B_{ij}\cos(\psi_i+\psi_j-\chi_{ij})\right];
\end{equation*}
the complex form makes the response to a common rotation and the field
exerted by the neighbours explicit. For positive \(B_{ij}\) the second term is
minimized by \(\psi_i=\chi_{ij}-\psi_j\pmod{2\pi}\): the preferred orientation
of one site is the reflection of its neighbour's. Rotating all orientations by
the same angle \(\delta\) leaves the first term unchanged but shifts the phase
of the second by \(2\delta\). Fixed bond directions therefore resist a
collective rotation, the motion through which a pattern held only by
alignment would drift away, even when the layer has little net orientation.

\begin{figure}[!htbp]
 \centering
 \includegraphics[width=\linewidth]{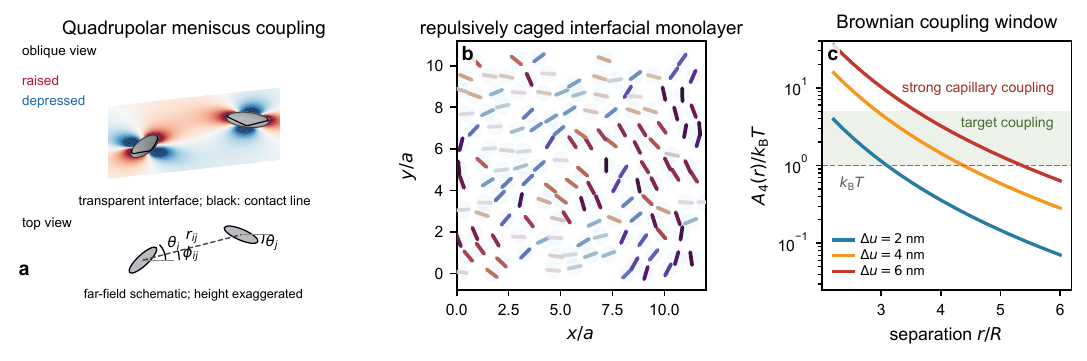}
 \caption{\label{fig:realization}
 \textbf{Capillary coupling in a caged layer of ellipsoids.}
 \textbf{a}, Raised (red) and depressed (blue) menisci around two
 ellipsoids. The top view defines the bond direction \(\phi_{ij}\) and the
 particle axes \(\theta_i,\theta_j\).
 \textbf{b}, A simulated layer at \((J,g)=(4,5)k_{\rm B}T\).
 Rod colour shows the orientation of the axis and pale discs the cage centres.
 \textbf{c}, Far-field capillary energy \(A_4\) against separation for
 interfacial tension \(\gamma=10\,\mathrm{mN\,m^{-1}}\) and three
 peak-to-trough heights \(\Delta u\) of the contact line. Shading marks
 \(1\)--\(5k_{\rm B}T\).}
\end{figure}

For ellipsoids with in-plane axes \(\theta_i\) at a liquid interface, the
second contribution comes from the meniscus (Fig.~\ref{fig:realization}a).
The contact line on an ellipsoid rises and falls around the particle, and for
an untilted particle with symmetric wetting and negligible vertical load the
leading far-field distortion is quadrupolar~\cite{loudet_capillary_2005,luo_self-assembly_2019}.
When the menisci of two particles overlap, the pair energy is proportional to
\(-\cos2(\theta_i+\theta_j-2\phi_{ij})\), where \(\phi_{ij}\) is the direction
of the line joining the centres. In Eq.~\eqref{eq:class} this corresponds to
\(\psi_i=2\theta_i\), \(\chi_{ij}=4\phi_{ij}\), \(A_{ij}=Jw^-_{ij}\) and
\(B_{ij}=gw^+_{ij}\), where \(J\) and \(g\) are the strengths of parallel
alignment and capillary coupling and the weights \(w^\mp_{ij}\) describe their
different ranges (Methods). At the capillary minimum the two axes are mirror
images across the bond, so a particle that turns changes the preferred
directions of all its neighbours. In the simulations the particle centres are
displaced randomly from a triangular lattice of spacing \(a\) by an amount
\(\sigma\) and kept in place, as in a dense layer in which electrostatic
repulsion cages the particles, while the axes undergo Brownian rotation with
diffusivity \(D_r\) (Fig.~\ref{fig:realization}b). A patterned field of
strength \(h\) per particle, as could be exerted by polarized light, writes a
chosen set of target angles.

\section*{A written pattern persists without net order}

A pattern of orientations is measured by comparing the axes of the
individual particles with a reference pattern. The connected overlap
subtracts the contribution of the mean orientations of the layer and of the
reference (Methods), so it distinguishes the persistence of a spatial pattern
from the persistence of a common director. The role of the capillary bonds
is clearest when the same angles are released into three versions of the
layer: one with capillary bonds, one with alignment only, and one in which the
capillary coefficient of every bond is added to its alignment coefficient, so
that each bond keeps the same total strength.

\begin{figure}[!htbp]
 \centering
 \includegraphics[width=\linewidth]{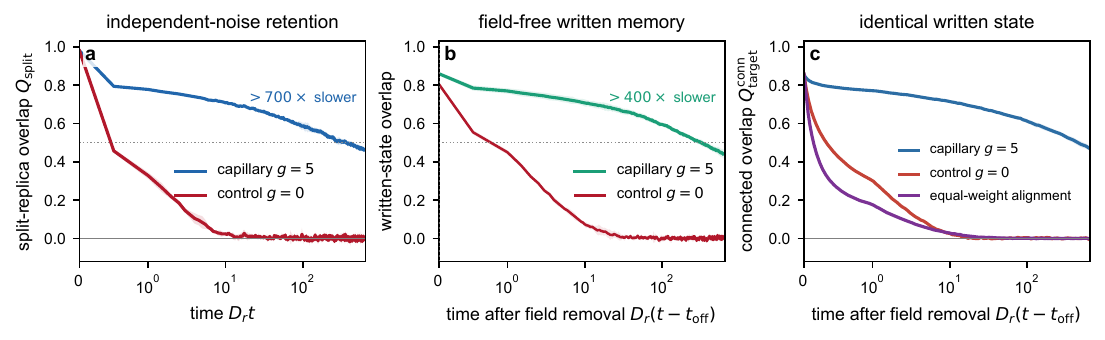}
 \caption{\label{fig:memory}
 \textbf{A pattern of orientations retained after the writing field is
 removed} (\(N=1024\)).
 \textbf{a}, Overlap of two copies of one configuration evolving with
 independent thermal noise. \textbf{b}, Overlap with a target written by a
 temporary field. Dotted lines mark overlap 0.5; the lower bounds on the ratio
 of the times to reach it with and without capillary coupling are 790 in
 \textbf{a} and 420 in \textbf{b}. Bands are standard deviations across
 three particle arrangements. \textbf{c}, Connected overlap with the target
 after release from identical angles at \(\sigma/a=0.16\) in the capillary
 layer and in the two alignment layers. Bands are standard errors across five
 arrangements.}
\end{figure}

Long after a free axis has forgotten its direction, the capillary layer
still keeps a connected overlap of \(0.47\)--\(0.50\) with the written
pattern, whereas both alignment layers lose it (Fig.~\ref{fig:memory}c). The
two losses differ. With the full bond strength in alignment the layer orders
almost completely along a common director, with nematic order
\(S=0.91\)--\(0.93\), yet even after the best rotation of that director its
overlap with the written pattern is only \(0.05\)--\(0.08\): alignment
converts the pattern into uniform order. The capillary bonds instead keep
each axis near its written direction while the net order stays small.
Measured by the time for the overlap to fall to one half, the capillary layer
keeps a written pattern at least 420 times longer than the layer without
capillary coupling, and two copies of one configuration evolving with
independent thermal noise remain alike at least 790 times longer
(Fig.~\ref{fig:memory}a,b), so the record is kept by the bonds and not by a
particular history of the noise.

The pattern is hidden from measurements that average over many particles.
Birefringence and polarized transmission respond to the net orientation
\(m=N^{-1}\sum_iz_i\), whose modulus is the nematic order \(S\). In the
capillary layer the written axes vary from particle to particle, their
contributions to \(m\) cancel, and a polarizer sees a nearly isotropic layer.
The pattern appears only when the axes of individual particles are compared
with the written ones. This is the opposite of the alignment layer, in which
the bonds turn the written pattern into a uniform director that a polarizer
detects at once but that no longer contains the pattern.

\section*{Frustrated loops create many metastable patterns}

\begin{figure}[!htbp]
 \centering
 \includegraphics[width=\linewidth]{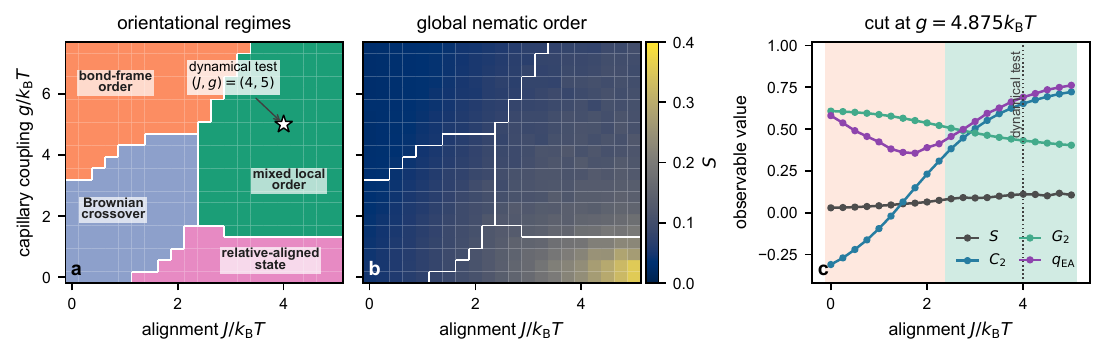}
 \caption{\label{fig:map}
 \textbf{Local and bulk order as the two interactions are varied.}
 \textbf{a}, Four classes of response from 441 coupling points on three
 layers of \(N=400\) particles. White curves separate the classes of the
 measured order parameters. \textbf{b}, Global nematic order. \textbf{c},
 Pair alignment and bond-frame order along the column \(g=4.875k_{\rm B}T\).
 The star marks \((J,g)=(4,5)k_{\rm B}T\), the point used in the other figures.}
\end{figure}

Whether a written pattern can survive without net order depends on a balance
between two kinds of local order. Parallel
alignment and the capillary preference can both be partly satisfied, and when
their strengths are comparable and exceed \(k_{\rm B}T\) the layer develops
strong local correlations of both kinds while keeping a weak director
(Fig.~\ref{fig:map}). Positional disorder is essential. On a regular
triangular lattice the fourth harmonics of the six symmetric bond directions
cancel for a uniform orientation, and a nearly regular layer is strongly
aligned, with \(S=0.92\) (Methods). Displacing the centres removes
this cancellation around each particle, and the preferred axes vary from site
to site, as in a random-anisotropy magnet~\cite{garanin_random_2022}. Unlike
in such a magnet, the local axes here are not imposed: they depend on the
orientations of the neighbours and move when the neighbours turn. Across the
disorder scan the global order falls thirteenfold while the local persistence
of the axes decreases only by a factor of 1.3
(Supplementary Section~S9).

The conflict between the bond preferences can be followed around a closed
loop of sites. Crossing a bond maps an orientation to its reflection,
\(T_{ij}(\psi)=\chi_{ij}-\psi\pmod{2\pi}\), and going around a loop
\(i_0,\ldots,i_{\ell-1}\) composes these maps. Around an even loop the result is
a rotation by the alternating sum of the bond phases \(\chi_{ij}\), which has a
fixed orientation only if the rotation vanishes; around an odd loop it is a
reflection, which has fixed orientations, but different loops through one
site generally fix different ones. No assignment of orientations satisfies
every bond, and the layer has many compromises of comparable energy. In a
reduced model in which each locally ordered region takes one of two
orientations, making the loops compatible at fixed bond strengths removes on
average \(5.6\) of the states that are stable against a single flip. These
frustrated loops supply the many metastable patterns in which records are
kept. How long a given pattern survives depends in addition
on its restoring torques and on the barriers that separate it from its
neighbours.

\section*{One layer stores many hidden patterns}

\begin{figure}[!htbp]
 \centering
 \includegraphics[width=0.92\linewidth]{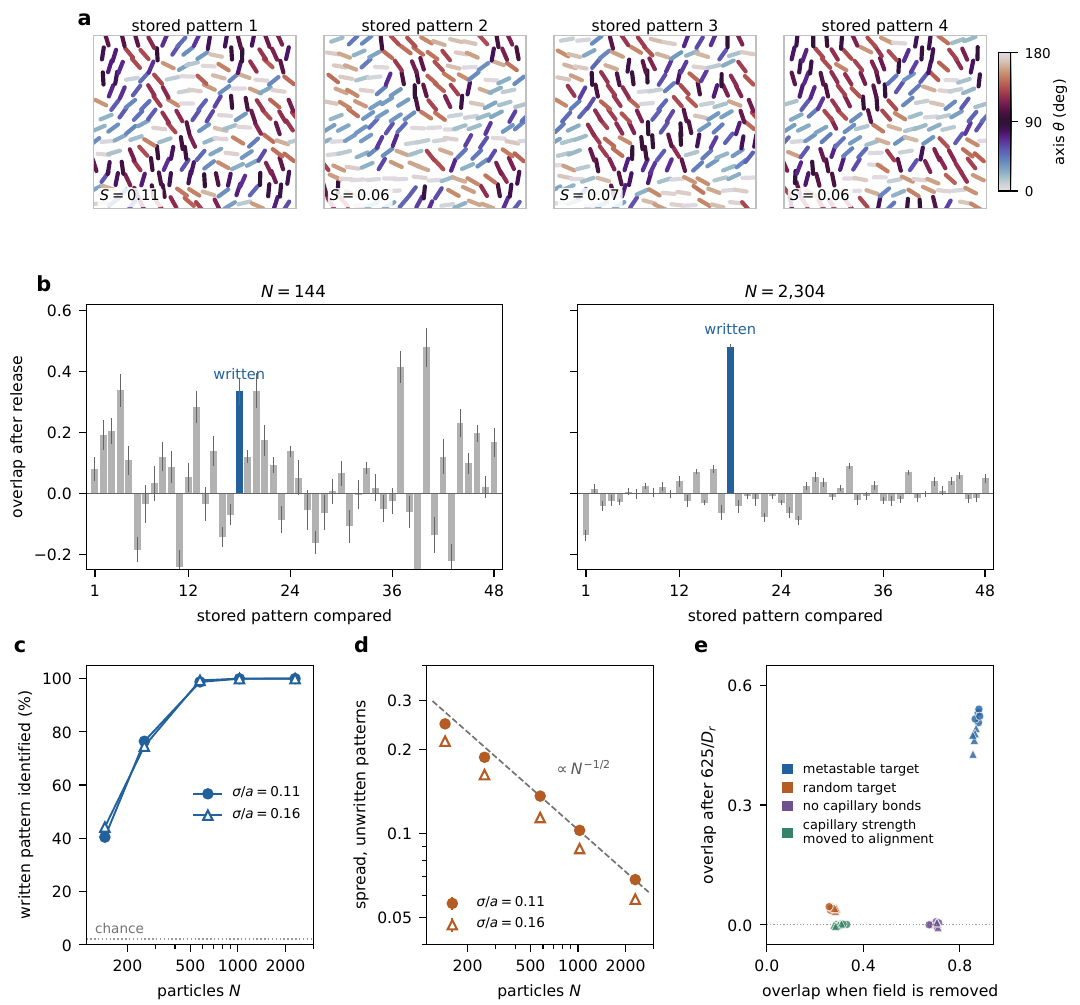}
 \caption{\label{fig:library}
 \textbf{Many hidden patterns stored in one arrangement of particles.}
 \textbf{a}, The same patch of one arrangement of \(N=2{,}304\) particles
 (\(\sigma/a=0.11\)) holding four of its 48 metastable patterns. The particle
 centres are identical in the four tiles; only the axes differ. Colour gives
 the axis angle, and \(S\) is the net order of the whole layer in each pattern.
 \textbf{b}, Connected overlap with each of the 48 stored patterns of the
 configuration retained \(625/D_r\) after pattern 18 was written, in a layer of
 144 particles and in one of 2,304 particles (\(\sigma/a=0.11\); mean and
 standard error over eight realizations). In the small layer other patterns
 overlap as strongly as the written one; in the large layer only the written
 pattern stands out.
 \textbf{c}, Fraction of final configurations in which the written pattern has
 the largest overlap among the 48; the dotted line is chance.
 \textbf{d}, Spread of the overlaps of single final configurations with the 47
 patterns that were not written; the dashed line has slope \(-1/2\). In
 \textbf{c} and \textbf{d}, filled and open symbols are \(\sigma/a=0.11\) and
 \(0.16\), and error bars are standard errors across five arrangements.
 \textbf{e}, Overlap with the target when the field is removed and after
 \(625/D_r\) for metastable and random targets in the capillary layer, and for
 metastable targets in the layer without capillary bonds and in the layer in
 which the capillary strength of every bond is moved to alignment
 (\(N=1024\)). Each point averages six targets and 16 realizations on one
 arrangement.}
\end{figure}

Because the loops admit many compromises, independent random initial angles
relaxed under the capillary interactions end in distinct metastable patterns
on a single arrangement of particle centres, and each of them can be
written. When such patterns are written one at a time by a temporary field
directed along them and the layer is then left without a field, it keeps a
mean connected overlap of \(0.46\)--\(0.51\) with the intended pattern while
the net order is only \(0.07\)--\(0.10\) (Methods). In every final
configuration the intended pattern has the largest overlap among all the
patterns written into the same layer, also when each alternative is given its
most favourable rotation of the layer
(Supplementary Fig.~S11).
Several such patterns occupy the same particles with little net order, so a polarizer cannot tell which of them the layer holds (Fig.~\ref{fig:library}a).

Patterns that the bonds do not stabilize cannot be stored. Targets made of
independent random angles are written to an overlap of only \(0.26\)--\(0.29\) by the same field and fall
to \(0.04\) after release, whereas in the layer without capillary coupling the
relaxed patterns are written well, to an overlap near \(0.7\), and then lost
(Fig.~\ref{fig:library}e). A pattern is retained when the field brings the
layer near a configuration that the bonds themselves stabilize. The usable
records of a layer are therefore its own metastable patterns, which are fixed
by the arrangement of the particles and can be measured once for each sample.

The number of patterns that one layer can tell apart grows with its size.
When 48 metastable patterns are written separately into layers of 144 to
2,304 particles, the retained overlap with the written pattern stays near
\(0.5\), whereas the overlap of the final configuration with each of the
other 47 patterns fluctuates about zero with a spread that falls
approximately as \(N^{-1/2}\), from \(0.25\) at \(N=144\) to \(0.07\) at
\(N=2{,}304\) (\(\sigma/a=0.11\); Fig.~\ref{fig:library}d). The written pattern is identified among
the 48 in \(40\%\) of final configurations at \(N=144\), in \(99\%\) at
\(N=576\) and in every configuration at \(N=2{,}304\) (Fig.~\ref{fig:library}b,c). The spread equals that
of about \(N/10\) independent orientations, as if the layer were composed of independent regions of about ten particles, comparable to the locally ordered regions that compete around frustrated loops. The ratio of the retained
overlap to this spread therefore grows as \(N^{1/2}\). If the overlaps with
rival patterns are independent and Gaussian, the number of patterns that can
be distinguished grows exponentially with \(N\), by about one binary choice
for every 50 particles, which corresponds to some 200 bits per write in a
layer of \(10^4\) particles (Methods).

\section*{Lowering the coupling erases and rewrites a pattern}

\begin{figure}[!htbp]
 \centering
 \includegraphics[width=\linewidth]{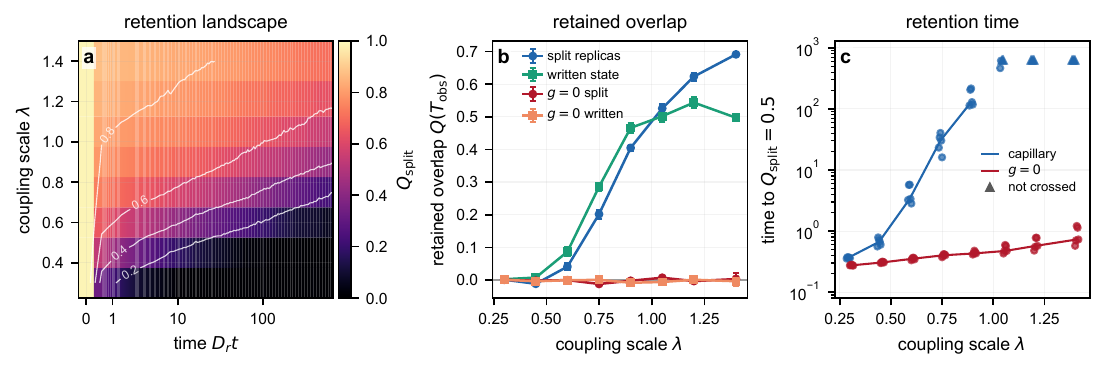}
 \caption{\label{fig:activated}
 \textbf{Retention as both interactions are strengthened.}
 \textbf{a}, Overlap of two copies of one configuration evolving with
 independent thermal noise for \((J,g)=\lambda(4,5)k_{\rm B}T\) on five
 layers of \(N=576\) particles. \textbf{b}, Overlap at \(625/D_r\) for such
 copies, for patterns written by a field and for layers without capillary
 coupling; error bars are standard errors across layers. \textbf{c}, Time at
 which the overlap of the copies first falls below 0.5. Triangles mark layers
 in which it stays above 0.5 throughout the observation.}
\end{figure}

Strengthening both interactions by a common factor \(\lambda\) raises the
energy barriers that keep these patterns, and the time over which two
copies of a configuration remain alike grows approximately exponentially with
the coupling, as expected for thermally activated escape
(Fig.~\ref{fig:activated}). The overlap with a pattern written by a fixed field
levels off instead: stronger bonds keep the layer longer in the state it
reaches, but they also resist the field, so the state reached resembles the
target less. The barriers that keep a memory therefore also oppose writing a
new one.

The two roles can be separated in time by changing the coupling during the
write. A weak field applied to a stored
state at fixed coupling leaves an overlap of only \(0.05\) with a new target.
When the coupling is lowered at the start of the write and restored gradually
before the field is removed, the same field leaves an overlap of \(0.56\) with
the new target, and the overlap with the previous state falls from \(0.59\) to
\(0.02\). The old pattern is erased and the new one written in one operation,
after which the restored barriers keep it. In a colloidal layer the coupling
could be varied in this way through the interfacial tension or the wetting of
the particles. Finite-size comparisons of independent layers show no
equilibrium spin-glass transition in the scanned range
(Supplementary Section~S12); the memory is a long-lived
nonequilibrium state, which is what writing and rewriting require.

\section*{The layer records the order of its writes}

When several fields act on the same particles, each new pattern is written
into a layer that already carries the earlier ones. Consider eight metastable
patterns, A to H, written in succession into a relaxed layer by brief
fields, after which the layer is left without a field (Methods). The final configuration can be
expressed as a combination of all the patterns involved,
including the starting state and two metastable patterns that were never
written. Because two metastable patterns of the same layer are not
orthogonal, the combination is obtained by solving with their full overlap
matrix (Methods); each coefficient measures how much of one pattern the final
configuration contains, and the coefficients of the two unwritten patterns
measure the fluctuations against which a trace must be detected.

\begin{figure}[!htbp]
 \centering
 \includegraphics[width=\linewidth]{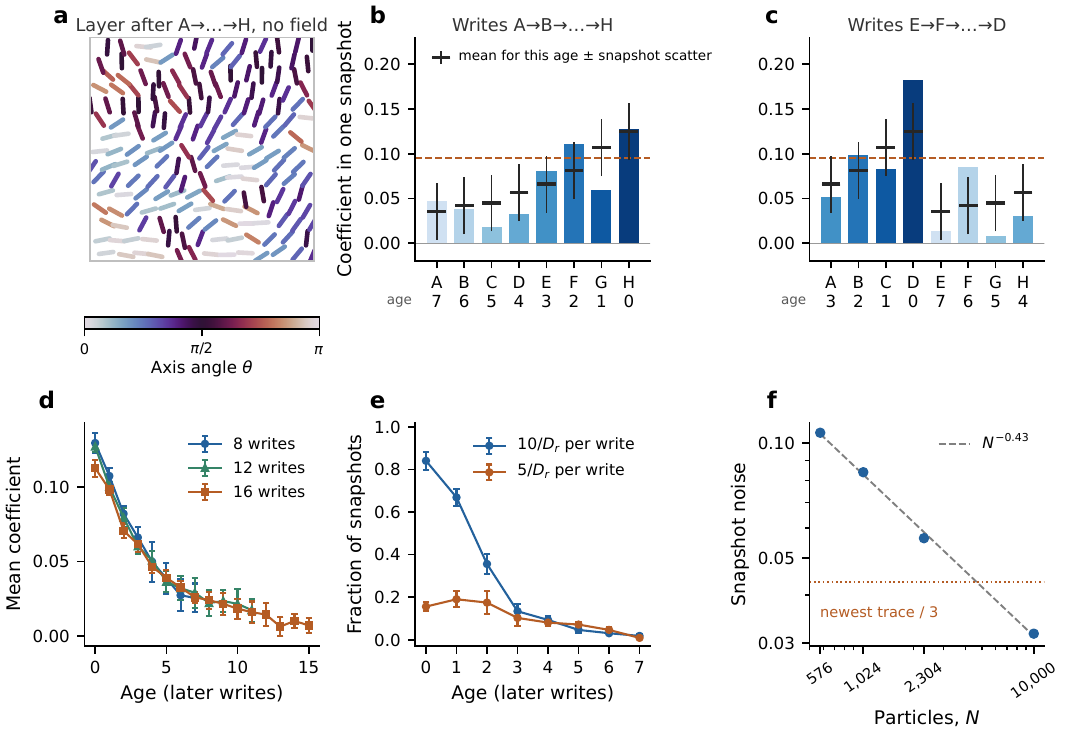}
 \caption{\label{fig:sequence}
 \textbf{Traces of successive writes in the orientations of one layer}
 (\(\sigma/a=0.11\), \(h=1\,k_{\rm B}T\)).
 \textbf{a}, Part of a single configuration of \(N=10{,}000\) particles
 \(625/D_r\) after eight patterns A--H were written in succession for
 \(10/D_r\) each; colour gives the axis angle.
 \textbf{b}, Coefficients of the eight written patterns in that configuration.
 The number under each letter is the age of the write, the number of writes
 that followed it. Black marks give the mean coefficient at each age over all
 320 configurations of this size, and vertical lines the fluctuation of a single
 configuration about it, so neighbouring older writes can exchange their order
 in one snapshot. The dashed line marks three times the fluctuation of the
 coefficients of patterns that were never written.
 \textbf{c}, The same for a configuration of the same layer written in the order
 E--D. Both configurations rank pairs of writes as well as the mean over all
 320 configurations.
 \textbf{d}, Mean coefficient against age for sequences of 8, 12 and 16 writes
 at \(N=1024\). \textbf{e}, Fraction of single configurations of
 \(N=10{,}000\) particles in which a write is detected above three times the
 fluctuation, for writes of \(10/D_r\) and \(5/D_r\); 320 configurations per
 duration from five arrangements and eight cyclic orders.
 \textbf{f}, Fluctuation of the unwritten coefficients against the number of
 particles, with a power-law fit; the dotted line is one third of the mean
 coefficient of the newest write. Error bars are standard errors across five
 arrangements.}
\end{figure}

A single configuration of 10,000 particles after the writes A to H shows no
visible order (Fig.~\ref{fig:sequence}a), yet its coefficients rise from the
oldest write to the newest (Fig.~\ref{fig:sequence}b). When the same eight
fields act in the order E to D, the profile moves with the order, and the
largest coefficient belongs to D (Fig.~\ref{fig:sequence}c). The layer
therefore records which patterns were applied and, in its graded amplitudes,
how recently each acted. Averaged over realizations, the coefficient falls
with the number of later writes by a factor of about 0.8 per write
(Fig.~\ref{fig:sequence}d). The profile depends on the age of a write and hardly on
the length of the sequence: after eight, twelve or sixteen writes the
coefficients at equal age nearly coincide, and only the newest write is
slightly weaker in the longest sequence. This is the defining
property of a palimpsest~\cite{nadal_palimpsests_1986,parisi_forgets_1986}.
Each new write reduces every earlier trace by a similar fraction, so the
record keeps the most recent writes regardless of how many preceded them.
In the layer in which the capillary strength of every bond is moved to alignment, all coefficients are below \(0.005\).

The graded profile follows from the way a trace is lost. When the last field
is switched off, the newest pattern has a coefficient of \(0.45\), which
relaxes to \(0.13\) during the field-free interval, while the coefficients of
the earlier writes change much less (Supplementary Fig.~S12).
At the end of a write the field holds the layer partly outside the
metastable states the bonds can stabilize, and on release the layer falls
into the nearest of them, keeping only the part of the newest pattern that
such a state contains. For the earlier patterns this loss has already taken
place: each subsequent field acts as their release, and what survives it is
carried into the final configuration. The fraction lost to each later write is
set by how far that write moves the layer. Halving the duration of each write halves the newest coefficient but lengthens the memory, so that
the coefficients decay over about ten writes instead of five
(Fig.~\ref{fig:sequence}e and Supplementary Fig.~S12).
The duration of the exposure thus trades the contrast of the most recent
traces against the depth of the record.

Whether the record can be seen in a single configuration, as in one image of
an experiment, depends on the number of particles. The mean coefficients do not change with the size of the layer, but their fluctuations decrease approximately as \(N^{-1/2}\), for the same reason as the spread of the overlaps with rival patterns, from \(0.107\) at \(N=576\) to \(0.032\) at
\(N=10{,}000\) (Fig.~\ref{fig:sequence}f). At the larger size the newest,
second and third newest writes exceed three times this fluctuation in
\(84\%\), \(67\%\) and \(36\%\) of single configurations, and at least two of
the eight writes do so in \(78\%\) of them. Ranking the coefficients of a
single configuration places two writes in the correct order of age in \(78\%\)
of pairs. The order of a whole sequence of eight is recovered only rarely,
because the oldest traces approach the fluctuation level, but the recent
history is legible in one snapshot. Two opposing fields give the limiting
case of a short sequence: when they act on shared particles, the field that
acted last is identified in every realization at the largest tested field,
compared with \(55\%\) when the two fields act on disjoint particles
(Supplementary Section~S16).

\section*{Dipolar layers obey the same interaction}

Equation~\eqref{eq:class} is not specific to capillarity. Two planar point
dipoles with angles \(\psi_i,\psi_j\) at separation \(r_{ij}\) interact with
energy
\begin{equation}
 U_{ij}^{\rm dip}=-\frac{C}{2r_{ij}^3}
 \left[\cos(\psi_i-\psi_j)+3\cos(\psi_i+\psi_j-2\phi_{ij})\right],
 \label{eq:dipoles}
\end{equation}
where \(C>0\) sets the electric or magnetic strength (Methods). The second
term again reflects the neighbouring orientation, with
\(\chi_{ij}=2\phi_{ij}\) for polar moments, and dominates the first. Dipoles
on fixed disordered sites are known to age~\cite{cortes-morales_equilibration_2016}.
In simulations of planar dipoles on disordered arrangements the mean net
moment is only \(0.04\)--\(0.05\), the time over which a prepared pattern
persists grows strongly with the dipolar coupling, and a weak patterned field
shortens that time by a factor of 22 (Methods). Magnetic nanoparticles or polar molecules on disordered sites,
and colloids carrying induced dipoles, are candidates for the same kind of
record.

\section*{A hidden and rewritable memory patch}

A sealed interfacial layer of elongated particles would be written by patterned,
polarized light and read only by comparison with its own metastable patterns,
which are measured once for each layer. An image that resolves the axes of
individual particles, as video microscopy of colloidal ellipsoids does,
compared with these patterns shows which of them the layer holds and, from the
graded coefficients, which were written most recently. The same image without
the patterns shows a disordered layer, and a polarizer shows an isotropic one.
Lowering the coupling during a write replaces one record by another. Because the patterns are fixed by a random arrangement of particles
that is difficult to reproduce, the layer resembles a physical one-way
function~\cite{pappu_physical_2002} that can in addition be rewritten and that
records the order of the challenges it receives. The depth of the history is
set by the duration of each exposure and its legibility by the number of
particles imaged: by the estimate from the spread of overlaps, a patch of
\(10^4\) particles, a few hundred micrometres across for micrometre-sized
ellipsoids, distinguishes about 200 bits in each write.

A patterned optical field can exert
torque on anisotropic particles through their polarizability~\cite{lapointe_towards_2011},
and interfacial curvature provides another means of turning them~\cite{cavallaro_curvature-driven_2011}.
Charged colloids can be caged in position while remaining free to
rotate~\cite{besseling_equilibrium_2026}, and Brownian rotation in dense
colloidal assemblies can be tracked directly~\cite{melio_pivoting_2026}.
For an interfacial tension of \(10\,\mathrm{mN\,m^{-1}}\) and a separation of
three effective contact-line radii, capillary bonds of \(1\)--\(5k_{\rm B}T\)
require a peak-to-trough height of the contact line of \(1.9\)--\(4.2\,\mathrm{nm}\);
porosity or surface roughness can reduce the much stronger capillary
attractions of smooth particles to this range~\cite{trevenen_nanoscale_2023,rahman_controlling_2026}.
The meniscus must relax faster than a particle turns, and the cages must last
for the duration of writing and imaging; in simulations with mobile centres,
the record survives root-mean-square displacements of about \(0.1a\). The
angular fluctuations of isolated pairs would calibrate the interaction before
the collective experiment, and tracking centres together with axes would
separate the loss of a record by cage drift from its loss by rotation.
Photoaligned surfaces, on which a prescribed angular pattern is written into a
fixed substrate~\cite{wu_arbitrary_2012}, provide a benchmark for the writing
step. In the colloidal layer, however, the pattern is carried by mobile
particles whose mutual torques select which patterns last and erase older
writes gradually as new ones arrive, so the material records a history and
not only its last state.

\section*{Methods}

\subsection*{Particle arrangement and interactions}

Particle centres are obtained by displacing the sites of a triangular
lattice of spacing \(a\) by independent Gaussian offsets of component
standard deviation \(\sigma\). The layer has periodic boundaries and
minimum-image bonds of length \(r_{ij}\leq2.6a\); pairs closer than \(0.55a\)
are omitted. The median nearest-neighbour spacing is \(r_0\). The capillary
weight is \(w^+_{ij}=\min[6,(r_0/r_{ij})^4]\), and the alignment weight
\(w^-_{ij}\) is \(\exp[-\max(r_{ij}/r_0-1,0)/0.2]\) for \(r_{ij}\leq1.35r_0\)
and zero otherwise. The capillary scale is \(g=A_4(r_0)\) with
\(A_4(r)=3\pi\gamma(\Delta u_{\max})^2(R/r)^4\), where \(R\) is the effective
contact-line radius and \(\Delta u_{\max}\) the peak-to-trough height of the
contact line. The meniscus is treated as quasistatic and the interactions as
pairwise additive. For the capillary layer the energy is
\(U=-\sum_{(ij)}[Jw^-_{ij}\cos2(\theta_i-\theta_j)+gw^+_{ij}\cos2(\theta_i+\theta_j-2\phi_{ij})]\).
With \(z_i=e^{2\mathrm{i}\theta_i}\), the neighbours of particle \(i\) exert
the field \(H_i=\sum_j[Jw^-_{ij}z_j+gw^+_{ij}e^{4\mathrm{i}\phi_{ij}}\bar z_j]\);
at fixed neighbours the \(\theta_i\)-dependent energy is
\(-\operatorname{Re}(\bar z_iH_i)\), minimized at \(\theta_i=\arg(H_i)/2\)
modulo \(\pi\). The curvature of the energy under a uniform rotation of all
axes is \(16gG_2\sum_{(ij)}w^+_{ij}\), with \(G_2\) defined below.

\subsection*{Rotational dynamics and writing}

For dimensionless time \(\tau=D_rt\) the axes obey
\begin{equation}
 d\theta_i=-\partial_{\theta_i}\left(\frac{U+U_{\rm write}}{k_{\rm B}T}\right)
 d\tau+\sqrt{2}\,dW_i,\qquad
 \langle dW_i dW_j\rangle=\delta_{ij}d\tau ,
\end{equation}
with independent Wiener increments \(dW_i\), integrated by the
Euler--Maruyama method with \(\Delta\tau=0.0025\) and angles reduced modulo
\(\pi\). A writing field adds
\(U_{\rm write}=-h\sum_i\cos2(\theta_i-\theta_i^{\rm target})\), where \(h\) is
the field energy per particle and \(\theta_i^{\rm target}\) the target axis.
Unless stated otherwise, \((J,g)=(4,5)k_{\rm B}T\). For the comparison from
identical angles (Fig.~\ref{fig:memory}c), a pattern was written with
\(h=1.5k_{\rm B}T\) for \(\tau=125\) and released for \(\tau=625\); the same
release angles were used in the layer with \(g=0\) and in the layer in which
the coefficient of \(\cos2(\theta_i-\theta_j)\) becomes
\(Jw^-_{ij}+gw^+_{ij}\). Means and standard errors are over five arrangements
per disorder, each averaged over 48 realizations.

\subsection*{Retention without a field}

The layers of \(N=1024\) particles are released for \(625/D_r\), about
2,500 times the correlation time \(1/4D_r\) of a free axis. The retained
overlap is \(0.47\) at \(\sigma/a=0.16\) and \(0.50\) at \(\sigma/a=0.11\).

\subsection*{Order parameters and overlaps}

With \(m=N^{-1}\sum_iz_i\), the global nematic order is \(S=|m|\). The pair
and bond-frame order parameters are
\begin{equation}
 C_2=\frac{\sum_{(ij)}w^-_{ij}\cos2(\theta_i-\theta_j)}{\sum_{(ij)}w^-_{ij}},\qquad
 G_2=\frac{\sum_{(ij)}w^+_{ij}\cos2(\theta_i+\theta_j-2\phi_{ij})}{\sum_{(ij)}w^+_{ij}}.
\end{equation}
For a pattern \(p\) with orientations \(z_i^{(p)}\) and mean \(m_p\), the
connected overlap is
\(Q_p^{\rm conn}=\operatorname{Re}[N^{-1}\sum_iz_i\bar z_i^{(p)}-m\bar m_p]\);
its modulus before taking the real part gives the largest connected overlap
under a common rotation. For two copies \(a,b\) of one configuration with
independent noise,
\(Q_{\rm split}=N^{-1}\sum_i\cos2(\theta_i^a-\theta_i^b)\). The persistence
\(q_{\rm EA}=N^{-1}\sum_i|\langle z_i\rangle_t|^2\) uses orientations averaged
over a finite time window.

\subsection*{Disorder scan}

The disorder scan varies \(\sigma\) from \(0.02a\), the nearly regular
layer, to \(0.28a\), beyond which particles would overlap, at \(N=576\) and
\((J,g)=(4,5)k_{\rm B}T\) (Supplementary Section~S9).

\subsection*{Separately written patterns}

For each of five arrangements of \(N=1024\) particles at each disorder,
24 metastable patterns were obtained by relaxing independent random angles for
\(125/D_r\). Each was written in a separate run from random angles with
\(h=1.5k_{\rm B}T\) for \(125/D_r\), with 16 realizations per pattern and
arrangement, and released for \(625/D_r\). A final configuration is assigned
to the pattern with the largest connected overlap. For the comparison under
rotation, the moduli of the complex connected overlaps are averaged over the
16 realizations for each written pattern and compared with those of its
alternatives. The comparisons with random targets and with the alignment
layers use six targets per arrangement.

\subsection*{Successive writes}

Each run starts from a metastable pattern of the arrangement that was allowed
to evolve again before writing. \(K=8\), 12 or 16 patterns are written back to
back with \(h=1\,k_{\rm B}T\) for \(10/D_r\) or \(5/D_r\) each, followed by
\(625/D_r\) without a field. All \(K\) cyclic orders of each sequence are used,
so that every pattern occupies every position once. Orientations are centred
by removing their mean, \(y_p\) denotes the centred vector of pattern \(p\)
and \(z\) that of the final configuration, and the coefficients \(c_p\) minimize
\(\|z-\sum_pc_py_p\|^2\) over the written patterns, the starting pattern and
two metastable patterns of the same arrangement that were never written. The
solution uses the full complex overlap matrix of these patterns, and the real
parts of \(c_p\) are reported. For \(N=1024\), five arrangements and 16
realizations per order were used at each disorder. For \(N=10{,}000\), eight
cyclic orders with eight realizations each on five arrangements give 320
configurations per write duration. A write is detected when its coefficient
exceeds three times the standard deviation of the coefficients of the two
unwritten patterns, pooled over all configurations of the same size and
duration. Relative age is inferred by ranking the eight coefficients of each
configuration.

\subsection*{Capacity against the size of the layer}

For \(N=144\), 256, 576, 1024 and 2304 at each disorder, five arrangements
were prepared. On each, 48 metastable patterns were obtained by relaxing
independent random angles for \(125/D_r\), and every pattern was written
separately from random angles with \(h=1.5k_{\rm B}T\) for \(125/D_r\) in eight
realizations and released for \(625/D_r\). The spread of the rival overlaps is
the standard deviation of the connected overlap of single final configurations
with the 47 patterns that were not written. Let \(m\) be the retained overlap,
less the mean rival overlap, divided by this spread. The largest of \(K-1\)
independent Gaussian rival overlaps reaches the retained overlap when
\(K\approx\exp(m^2/2)\). For \(N\geq576\), \(m^2/N=0.025\) at \(\sigma/a=0.11\)
and \(0.028\) at \(\sigma/a=0.16\), which gives \(\log_2K\approx N/55\) and
\(N/50\). The fitted exponents of the spread against \(N\) are \(-0.46\) at
both disorders.

\subsection*{Rewriting at variable coupling}

Five arrangements of \(N=576\) particles at \(\sigma/a=0.16\) were prepared in
relaxed states. A field of \(0.25k_{\rm B}T\) per particle was applied for
\(100/D_r\) towards an independently relaxed target, either at fixed coupling
\((J,g)=1.05(4,5)k_{\rm B}T\) or while the factor multiplying \((4,5)k_{\rm B}T\)
rose from \(0.30\) to \(1.05\) in 40 steps. Both were then evolved without a
field at the higher coupling for \(200/D_r\), and overlaps were averaged over
the last quarter of this interval.
The field of \(0.25k_{\rm B}T\) per particle is about a twentieth of the
capillary coupling.

\subsection*{Planar dipoles}

The angular factor of the energy of two equal planar point dipoles,
\(\cos(\psi_i-\psi_j)-3\cos(\psi_i-\phi_{ij})\cos(\psi_j-\phi_{ij})\), expands to
Eq.~\eqref{eq:dipoles}, with \(C=\mu_0\mu^2/4\pi\) for magnetic moments \(\mu\)
in free space and \(C=p^2/4\pi\epsilon\) for electric moments \(p\) in a medium
of permittivity \(\epsilon\). The simulations use five disordered arrangements
of \(N=576\) sites at \(\sigma/a=0.16\) with \(r^{-3}\) weights and a cutoff of
\(3.5a\). The time for leaving a prepared pattern is measured as the first fall
of its overlap below half of its short-time value in the initial state, and
the quoted factor compares geometric means over the five arrangements with and
without the patterned field.
The factor of 22 is obtained at \(C/(r_0^3k_{\rm B}T)=3\) with a patterned
field of \(0.75k_{\rm B}T\) per dipole.

\subsection*{Computational assistance}

The Hyperion system helped formulate the bond-direction mechanism and the
comparisons used to test it. OpenAI Codex and Anthropic Claude assisted with
simulation code, figure preparation, numerical checks and editing of the text.
The author selected the model, interpreted the calculations and checked the
results.

\section*{Data availability}

All simulation records are available on Zenodo
(\href{https://doi.org/10.5281/zenodo.24694114}{doi:10.5281/zenodo.24694114}),
including the separately written patterns, the capacity against layer size,
the successive writes, the rewriting at variable coupling and the dipole
arrays.

\section*{Code availability}

The simulation and analysis code is available at
\url{https://github.com/vbaulin/geometry-encoded-orientational-memory}.

\bibliography{memory,rotating_colloids_nature_physics_refs,rotating_colloids_writable_materials_refs,nature_physics_additions,nature_physics_reframed}
\clearpage
\includepdf[pages=-]{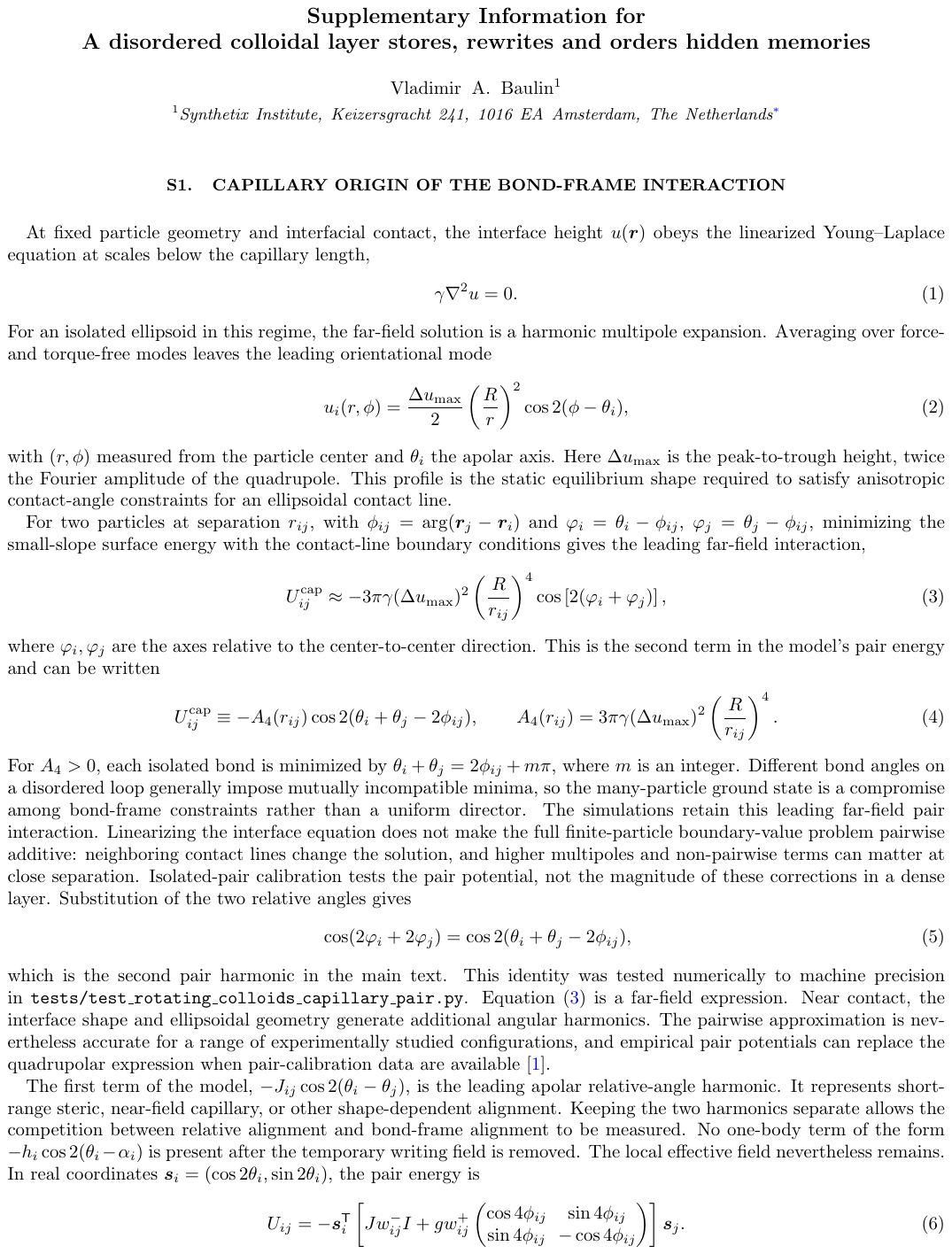}
\end{document}